\documentclass[superscriptaddress,aps,prl,amsmath,amssymb,floatfix,reprint,nobalancelastpage,raggedbottom,longbibliography]{revtex4-1}
\usepackage{graphicx}
\usepackage[colorlinks=true,citecolor=blue,linkcolor=blue,urlcolor=blue]{hyperref}
\usepackage{amsmath}
\usepackage{dcolumn}
\usepackage{xcolor}
\usepackage{fancyhdr}
\usepackage{lipsum}
\usepackage{soul}
\usepackage{balance}

\usepackage{footnote}

\makeatletter 
\renewcommand{\fnum@figure}{\textbf{Fig.~\thefigure}}
\makeatother

\makeatletter
\def\bbordermatrix#1{\begingroup \m@th
  \@tempdima 4.75\p@
  \setbox\z@\vbox{%
    \def\cr{\crcr\noalign{\kern2\p@\global\let\cr\endline}}%
    \ialign{$##$\hfil\kern2\p@\kern\@tempdima&\thinspace\hfil$##$\hfil
      &&\quad\hfil$##$\hfil\crcr
      \omit\strut\hfil\crcr\noalign{\kern-\baselineskip}%
      #1\crcr\omit\strut\cr}}%
  \setbox\tw@\vbox{\unvcopy\z@\global\setbox\@ne\lastbox}%
  \setbox\tw@\hbox{\unhbox\@ne\unskip\global\setbox\@ne\lastbox}%
  \setbox\tw@\hbox{$\kern\wd\@ne\kern-\@tempdima\left[\kern-\wd\@ne
    \global\setbox\@ne\vbox{\box\@ne\kern2\p@}%
    \vcenter{\kern-\ht\@ne\unvbox\z@\kern-\baselineskip}\,\right]$}%
  \null\;\vbox{\kern\ht\@ne\box\tw@}\endgroup}
\makeatother

\begin{document}
\title{Equivalent Circuit for Magnetoelectric Read and Write Operations}
\author{Kerem Y.  Camsari}
\affiliation{School of Electrical and Computer Engineering, Purdue University, IN, 47907}
\author{Rafatul Faria}
\affiliation{School of Electrical and Computer Engineering, Purdue University, IN, 47907}
\author{Orchi Hassan}
\affiliation{School of Electrical and Computer Engineering, Purdue University, IN, 47907}
\author{Brian M. Sutton}
\affiliation{School of Electrical and Computer Engineering, Purdue University, IN, 47907}
\author{Supriyo Datta}
\affiliation{School of Electrical and Computer Engineering, Purdue University, IN, 47907}
\date{\today}

\begin{abstract}

We describe an equivalent circuit model applicable to a wide variety of magnetoelectric phenomena and use SPICE simulations to benchmark this model against experimental data. We use this model to suggest a different mode of operation where the ``1'' and ``0'' states are not represented by states with net magnetization (like $m_x$, $m_y$ or $m_z$) but by different easy axes, quantitatively described by ($m_x^2 - m_y^2$) which switches from ``0'' to ``1'' through the write voltage. This change is  directly detected as a read signal through the inverse effect. The use of ($m_x^2 - m_y^2$) to represent a bit is a radical departure from the standard convention of using the magnetization ($m$) to represent information. We then show how the equivalent circuit can be used to build a device exhibiting tunable randomness and suggest possibilities for extending it to non-volatile memory with read and write capabilities, without the use of external magnetic fields or magnetic tunnel junctions. \end{abstract}
\maketitle

\section{INTRODUCTION}
In magnetic random access memory (MRAM) technology write units are typically based on spin-torque or spin-orbit torque, while read operations are based on the magnetoresistance of magnetic tunnel junctions (MTJ). But there is increasing interest in voltage-driven units due to the potential for low power operation, both active and stand-by based on different types of magnetoelectric phenomena \cite{biswas2017experimental, roy2011hybrid, kani2017strain, jaiswal2017mesl, manipatruni2015spin, gao2017electric, heron2014deterministic,he2010robust, zhao2016magnetoelectric, amiri2012voltage,chien2016enhanced, piotrowski2016size, mankalale2017comet, khan2014voltage, pertsev2008giant, SONG201733, peng2017speed,PhysRevB.80.224416,iraei2017proposal, sharmin2017magnetoelectric}.

\begin{figure}[t!]
\includegraphics[width=1\linewidth]{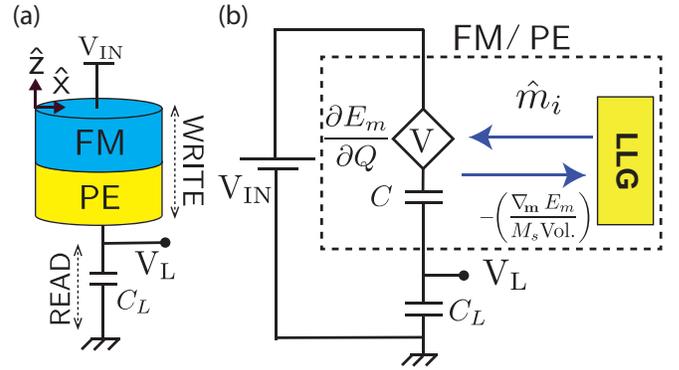}
\caption{\textbf{Equivalent circuit  for magnetoelectric (ME) read and write operations} (a) The charge on the piezoelectric (PE) capacitor changes the easy-axis of the ferromagnet (FM) and this causes a change in the output voltage $\rm V_L$ through the inverse effect. (b) Equivalent circuit model obtained from (\ref{eq:eqA}).  Write operation is through the effective field $\vec{H}_{me}=-\nabla_\mathbf{m} \ E_m /(M_s \mathrm{Vol.}$) that enters the stochastic Landau-Lifshitz-Gilbert (s-LLG) equation. Read operation is through the dependent voltage source V that is proportional to  $\partial E_m / \partial Q$, where $E_m$ is the magnetic energy.}
\label{fi:fig1}
\end{figure}

The central result of this paper is an equivalent circuit model (Fig.~\ref{fi:fig1}) applicable to a range of magnetoelectric (ME) phenomena including both write and read operations. It consists of a capacitor circuit which incorporates the back voltage from the magnetoelectric coupling described by (\ref{eq:eqA}):     	
\begin{equation}
\mathrm{V_{\rm IN}} = \displaystyle\frac{Q}{C_L}+\frac{Q}{C}+\displaystyle\frac{\partial E_m}{\partial Q}\
 \label{eq:eqA}
\end{equation}
where $E_m$ is the magnetic energy including the part controlled by the charge $Q$ on an adjacent capacitor $C$, through the ME effect. Equation (\ref{eq:eqA}) is solved self-consistently with the stochastic Landau-Lifshitz-Gilbert (s-LLG) equation which feels an effective field ($\vec{H}_{me}=-\nabla_\textbf{m} \  E_m/ \{M_s \mathrm{Vol.}\}$), $\nabla_\textbf{m}$ represents the gradient operator with respect to magnetization directions $\hat{m}_i$,  $M_s$ is the saturation magnetization and Vol. is the volume of the magnet. The s-LLG treatment for all simulations in this paper is similar to what is described in \cite{sun2000spin,sun2004spin,camsari2015modular} and is not repeated here. 
We first benchmark this equivalent circuit against the recently demonstrated MagnetoELectric Random Access Memory (MELRAM) device \cite{tiercelin2011room,klimov2017magnetoelectric} which uses the magnetoelectric effect (ME) and its inverse (IME) for write and read operations, {using a structure whose energy $E_m$ is given by Eq.~\ref{eq:eq3}.} We then argue that, unlike MELRAM, the ``1'' and the ``0'' states need not be represented by states with a net magnetization. {For example, using a structure whose energy $E_m$ is given by Eq.~\ref{eq:eq1}, one could instead switch the easy axis with a write voltage, and this change in the easy axis can be read} as a change in the voltage across a series capacitor through the inverse effect, allowing a ``field-free'' operation without any symmetry breaking magnetic field.

\section{EXPERIMENTAL BENCHMARK}
\begin{figure*}
\includegraphics[width=0.95\linewidth]{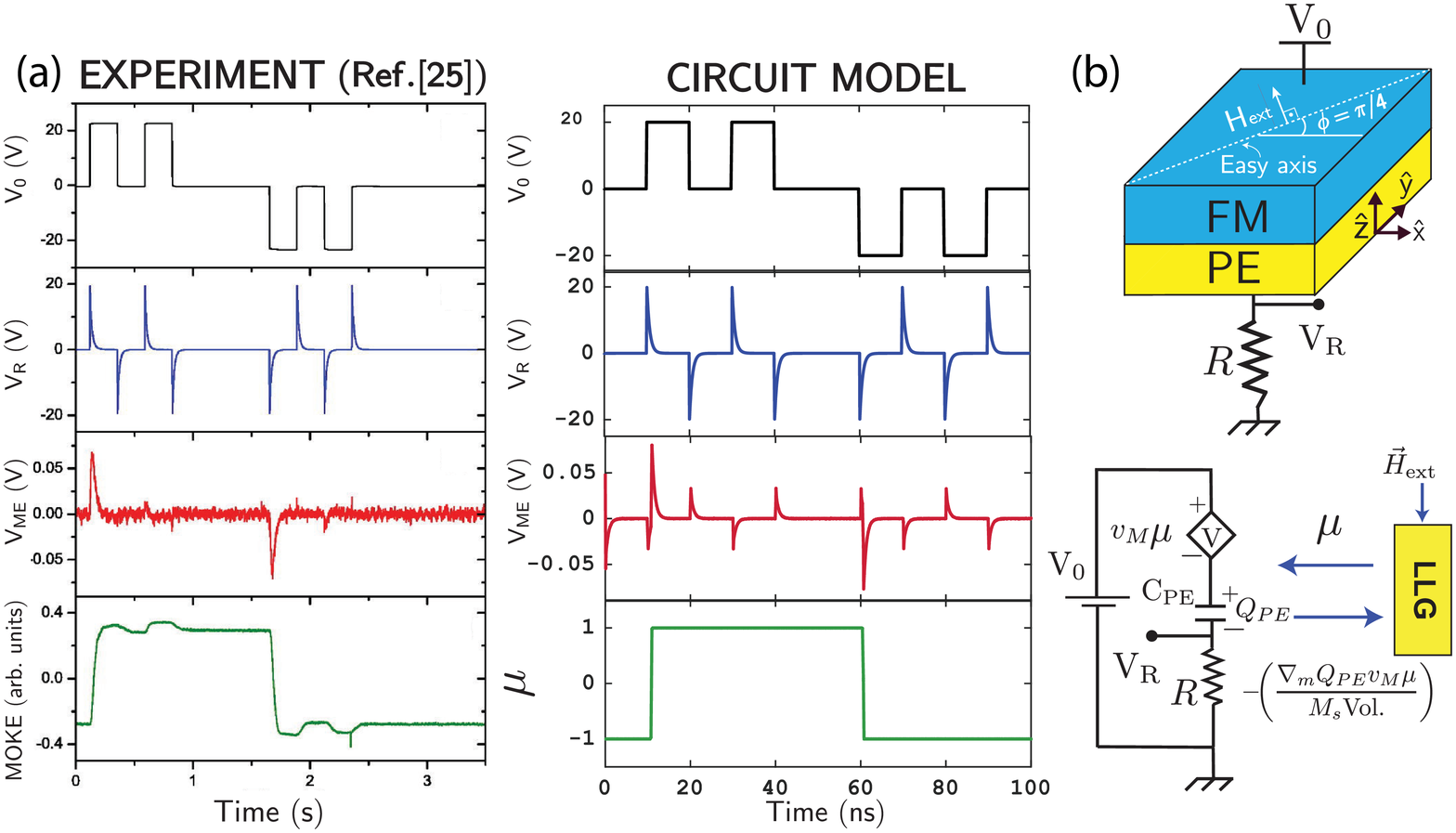}
\caption{\textbf{Experiment vs circuit model}: (a) The results of the self-consistent circuit model for the structure in (b) are in good agreement with the experimental results in \cite{klimov2017magnetoelectric}. $V_{\rm ME}$ is the mathematical difference of two measurements of $V_R$ with and without the external magnetic field, $V_{\rm ME}=V_R(H\neq0) -V_R(H=0)$. (b) {Experimental structure} reported in \cite{klimov2017magnetoelectric} where the piezeoelectric (PE) is $\langle 011\rangle$-cut PMN-PT and the ferromagnet (FM) is N layers of $\rm TbCo_2/FeCo$. The back-voltage is V=$v_M \mu$ where $\mu=m_x^2-m_y^2$ and the magnetic energy is $E_m= Q_{PE} v_M \mu $ where $Q_{PE}$ is the charge on the capacitor $\rm C_{PE}$. The following parameters are used: Coercivity for FM ($H_K$=200 Oe), saturation magnetization $M_s$=1100 emu/cc,  FM thickness, $t_{\rm FM}$=200 nm, PE thickness $t_{\rm PE}$=30 $\mu$m, Area=$520\times 520 \ \rm nm^2$, Magnetoelastic constant $B=-7$ MPa, a net PE constant, $d=d_{31}-d_{32}=2500$ pC/N, permittivity $\epsilon=4033 \ \epsilon_0$, resistance $R=\mathrm{2 \ M}\Omega$, back voltage  $v_M=B d t_{\mathrm{FM}}/2\epsilon$. In the experiment, magneto-optic Kerr effect (M.O.K.E) is used to show the variation of magnetization, which is compared to the pseudo-magnetization in our simulation. {Experimental panel is reproduced with permission of AIP Publishing LLC, from Reference \cite{klimov2017magnetoelectric}}.}
\label{fi:fig2}
\end{figure*}

We start with the MELRAM device (Fig.~\ref{fi:fig2}b) reported recently in \cite{klimov2017magnetoelectric} where the magnetic energy has the form
\begin{eqnarray}
&& E_m = -E_A m_x m_y  \nonumber \\
 && +E_H/\sqrt 2 (m_x -m_y) + v_M Q \  (m_x^2 - m_y^2) 
 \label{eq:eq3}
\end{eqnarray}
We note that this energy expression is essentially the same as what was reported in Ref.~\cite{klimov2017magnetoelectric} expressed using magnetization components, $m_x, m_y, m_z$. {For example, the anisotropy energy is written in ~\cite{klimov2017magnetoelectric} as $-E_A \sin^2 \phi$, with $\phi$ measured from the magnetic field $\vec{H}_{ext}$ such that $m_x=\cos(3\pi/4 - \phi), m_y=\sin(\pi/4 - \phi)$ and $m_x m_y = \sin^2 \phi$, ignoring an unimportant constant. Similarly the Zeeman term is written in ~\cite{klimov2017magnetoelectric} as $-E_H\cos\phi$ which equals $E_H (m_x-m_y)/ \sqrt{2}$. In ~\cite{klimov2017magnetoelectric},} the uniaxial anisotropy  energy term  and the external magnetic field were ingeniously balanced (by choosing $E_H= E_A \sqrt 2)$ to provide two unique low energy states that represent ``0" and ``1" at $\phi=\pi/2$ and $\phi=\pi$.

Finally, the last term represents the ME effect where an applied voltage generates a charge $Q$, controlled by the input voltage $V_{\rm IN}$, which changes the anisotropy energy such that  a positive (or negative) $Q$ causes the magnetic energy to favor the y-axis (or the x-axis) for  a positive $v_M$. This is  due to the anisotropic  piezoelectric coefficients $d_{31}$ and $d_{32}$ having different signs, a special property of the $\langle 011\rangle$-cut (PMN-PT)  that was chosen in the experiment. 

The equivalent circuit incorporates the back voltage from the ME coupling using  (\ref{eq:eqA}), with the load capacitor $C_L$ replaced by a resistor $R$: 
\begin{eqnarray}
\mathrm{V_{\rm IN}} =& \displaystyle R \frac{d Q}{dt}+\frac{Q}{C}+\displaystyle\frac{\partial E_m}{\partial Q}\nonumber \\
=& \displaystyle  \  \ \ \ R \frac{d Q}{dt} + \frac{Q}{C}+ v_M (m_x^2-m_y^2)
\label{eq:eq4}
\end{eqnarray}

\noindent {It is possible to write the ME energy as $q_M V$ in terms of an applied voltage V rather than charge $Q$, but this choice would lead to a back charge $\partial E_m / \partial V$ instead of a back voltage $\partial E_m / \partial Q$, giving a different but equivalent looking circuit model.}

Fig.~\ref{fi:fig2}a shows the write and read signals for the experimental structure in Fig.~\ref{fi:fig2}b calculated using a SPICE model, that are in good agreement with the experimental results presented in \cite{klimov2017magnetoelectric}. The reason for the very different time scales of the experiment and the circuit model is that the circuit model solves the real-time dynamics of the nanomagnet with time steps of the order of a fraction of the inverse FMR frequency of the nanomagnet ($1/f \sim 2\pi / \gamma / \sqrt{[H_K ( H_K +  4\pi M_s)]} \sim$ 0.2 ns for the chosen parameters) to avoid large numerical integrations while the experimental measurement is performed with quasi-static pulses. Therefore the RC time constants in both cases are very different, however  the maxima and minima  of each signal closely match based on the chosen parameters.

\section{FIELD-FREE OPERATION}

It is evident from Fig.~\ref{fi:fig2}a that our equivalent circuit describes the switching process accurately in the experiment described in Ref.~\cite{klimov2017magnetoelectric}. Using the same circuit model we would like to suggest the possibility of field-free operation where ``0" and ``1" are represented by two different easy axes rather than two different magnetization directions. { For this illustration, we consider a ferromagnet whose easy axis does not lie along $m_x=\pm m_y$ as in Ref.~\cite{klimov2017magnetoelectric}, but rather along the y-axis ($m_x=0$), corresponding to an anisotropy energy given by $E_A(m_x^2-m_y^2).$ Also, there is no external magnetic field so that $E_H=0$ giving an overall energy expression of the form}
\begin{equation}
E_m  = (E_A + v_M Q){(m_x^2 - m_y^2)}
\label{eq:eq1}
\end{equation}

\noindent {instead of Eq.~\ref{eq:eq3}.}

A positive induced charge $Q$ makes $y$-direction the easy axis so that $\langle m_x^2 - m_y^2\rangle= -1$, while a negative $Q$ makes $x$-direction the easy axis so that $\langle m_x^2 - m_y^2\rangle = +1$, and this constitutes the writing operation. The inverse of the same effect gives rise to a back voltage that allows one to read the information. Using (\ref{eq:eq1}) we obtain from (\ref{eq:eqA}):

\begin{equation}
\mathrm{V_{\rm IN}} =  \displaystyle{\frac{Q}{C_L}}+ \frac{Q}{C}+v_M {(m_x^2-m_y^2)}  
                \label{eq:eqB}
\end{equation}

Use of this ``pseudo-magnetization'' $ \mu \equiv m_x^2 - m_y^2$ is a radical departure from the standard convention of using the magnetization ($m_x, m_y$ or $m_z$) to represent a bit \cite{nikonov2017patterns}, opening up new possibilities for writing and reading.

Even though we have limited our discussion to the composite PE / FM structures that give rise to a magnetoelectric effect due to a coupling of the strain from a PE material to a magnetostrictive FM material, we believe the circuit description described in Fig.~{\ref{fi:fig1}} could be of general use. Indeed, it should be possible to use other quantities represented by a function $f(m_x,m_y,m_z)$ to represent a bit. Any mediating term due to strain, charge, orbital or  other microscopic mechanisms 
giving rise to a term of the form ($Q \times f $) in the energy expression that can be used to write such a bit, should also give rise to an inverse effect for read out. 

In the next two sections, we show two example uses of the external magnetic field-free operation of the equivalent circuit.

\begin{figure}[t!]
\includegraphics[width=0.90\linewidth]{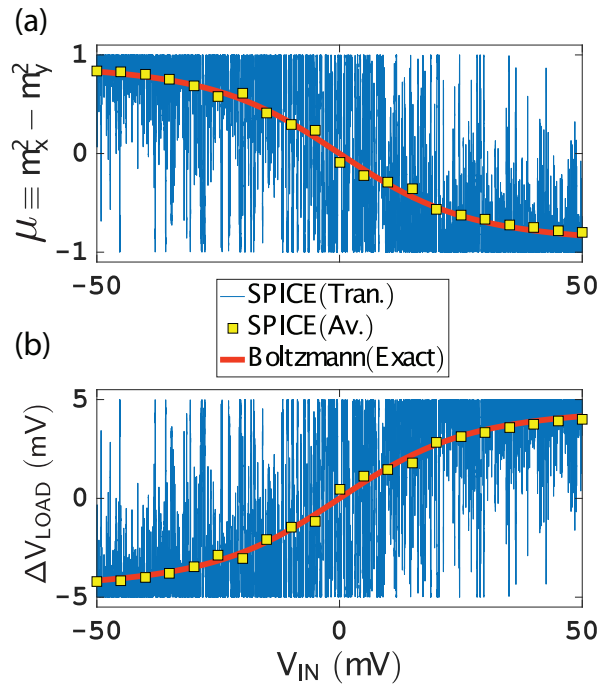}
\caption{\textbf{Tunable randomness:} Results for the structure in Fig.~\ref{fi:fig1}a using the circuit model in Fig.~\ref{fi:fig1}b with a circular magnet ($H_K\rightarrow 0, E_A\rightarrow 0$). $C=C_L=50 \rm \ aF$ and $v_M=10$ mV such that $C_{\rm eff} v_M^2 /kT <1$ (a) Three results are shown for magnetization, $\mu$ : Transient SPICE simulations (Solid blue) where the input voltage ($\rm V_{\rm IN}$) is swept from $-50$ mV to $+50$ mV in 1 $\mu$s and pseudo-magnetization, $\mu$ is plotted against $\rm V_{\rm IN}$. Separate SPICE simulations for each solid square where an average magnetization is obtained over 100 ns. Exact Boltzmann integral obtained from Eq.~\ref{eq:blaw}. (b) Same results for the differential load voltage, $  \Delta \mathrm{V_{L} = V_L - V_L}({v_M}=0)$, in this case $\mathrm{V_L}(v_M=0)= \rm V_{IN}/2.$ The actual load voltage has a linear $\rm V_{IN}$ dependence superimposed on $\Delta \rm V_L$, similar to Fig.~\ref{fi:fig5}. The differential load voltage is shown here for clarity.}
\label{fi:fig4}
\end{figure}

\section{Example \#1:  Tunable randomness}
The first example we illustrate using the equivalent circuit of Fig.~\ref{fi:fig1} is obtained by coupling the circuit shown in Fig.~\ref{fi:fig1} with a low-barrier circular nanomagnet that does not have an easy axis ($H_K\rightarrow 0)$  and  no energy barrier ($E_A=0)$ that favors a magnetization axis \cite{cowburn1999single,debashis2016experimental}. The magnetization of such a magnet fluctuates randomly in the plane, in the presence of thermal noise. The read and write mechanisms of the ME effect convert the fluctuations in the pseudo-magnetization $\mu$ to a voltage.

 Fig.~\ref{fi:fig4} shows the differential load voltage $\rm \Delta V_L$ vs $\rm V_{\rm IN}$ assuming $C=C_L=$50 aF  making $C_{\rm eff}=C C_L / (C+C_L) =25 \rm \ aF$ and $v_M$ = 10 mV, consistent with the material parameters for the experimental system in Fig.~\ref{fi:fig2}b, though the coupling coefficient $v_M$ is chosen somewhat smaller, (such that $C_{\rm eff} v_M^2 /kT <1$, as we explain in the next section) in order to avoid any hysteresis or memory effects. Alternatively one could use a smaller load capacitance, reducing $C_{\rm eff}$.

 With this choice of parameters, the magnetizations and hence the voltages fluctuate with time, and the averaged values over a time interval of $\approx$100 ns match the average results obtained from the Boltzmann probability:
\begin{equation}
\langle \mu \rangle =  \displaystyle\frac{\displaystyle\int_{Q=-\infty}^{Q=+\infty} \int_{\phi=-\pi}^{\phi=+\pi} d\phi \ dQ \ \overbrace{\cos (2\phi)}^{\mu=m_x^2-m_y^2} \rho(Q,\phi)}{\displaystyle\int_{Q=-\infty}^{Q=+\infty} \int_{\phi=-\pi}^{\phi=+\pi} d\phi  \ dQ \  \rho(Q,\phi)}
\label{eq:blaw}
\end{equation}
where $\rho (Q,\phi)= {1}/Z \ {\exp[-E(\phi,Q)}/{kT}]$ and $E = Q v_M \mu + Q^2/(2C_{\rm eff}) - Q \rm V_{\rm IN}$ represents the total energy. Similar to our previous discussion, we assume that the magnetization for the circular in-plane magnet is confined to the plane of the magnet  due to the strong demagnetization field. Therefore, the magnetization integral can be taken in the plane ($\phi \rightarrow \pm \pi$) and this seems to be in good agreement with the numerical s-LLG results as shown in Fig.~\ref{fi:fig4}. The average load voltage is obtained using Eq.~\ref{eq:blaw}, but replacing cos($2\phi)$ with $Q/C_L$.  

Eq.~\ref{eq:blaw} does not seem to reduce to a compact closed form, but assuming $Q  = C_{\rm eff} (V_{\rm IN}  \pm v_M |\mu|) \approx C_{\rm eff} V_{\rm IN}$ for small $v_M$, allows a direct evaluation:
\begin{equation}
\langle \mu \rangle \approx  - \frac{I_1 (x)}{I_0(x)}
\label{eq:bessel}
\end{equation}
where $I_n$ is the modified Bessel function of the first kind \cite{weisstein2006modified}, and $x= {Q v_M}/{kT}$. This approximation (not shown) seems to be in good agreement with an exact numerical evaluation of Eq.~\ref{eq:blaw} that is shown in Fig.~\ref{fi:fig4} and could be useful as an analytical guide. 

Note that the SPICE simulation solves the magnetization and the load voltage self-consistently following the equivalent circuit  (Fig.~\ref{fi:fig1}a) while the Boltzmann law takes these self-consistencies into account exactly. The agreement between the two constitutes another important benchmark for our equivalent circuit. 

With additional gain and isolation that can be incorporated by CMOS components this field-free voltage-tunable randomness can become a potential voltage controllable ``p-bit'' (probabilistic bit) that can be used as a building block for  a new type of probabilistic logic \cite{behin2016building,faria2017low, camsari2017stochastic,sutton2017intrinsic,pervaiz2017hardware,camsari2017p} or other neuromorphic approaches that make use of stochastic units \cite{suri2013bio,yu2013stochastic,suri2012cbram,zhang2015energy,srinivasan2016magnetic,sengupta2016magnetic}, but this is not discussed further. 

\begin{figure}[t!]
\includegraphics[width=0.88\linewidth]{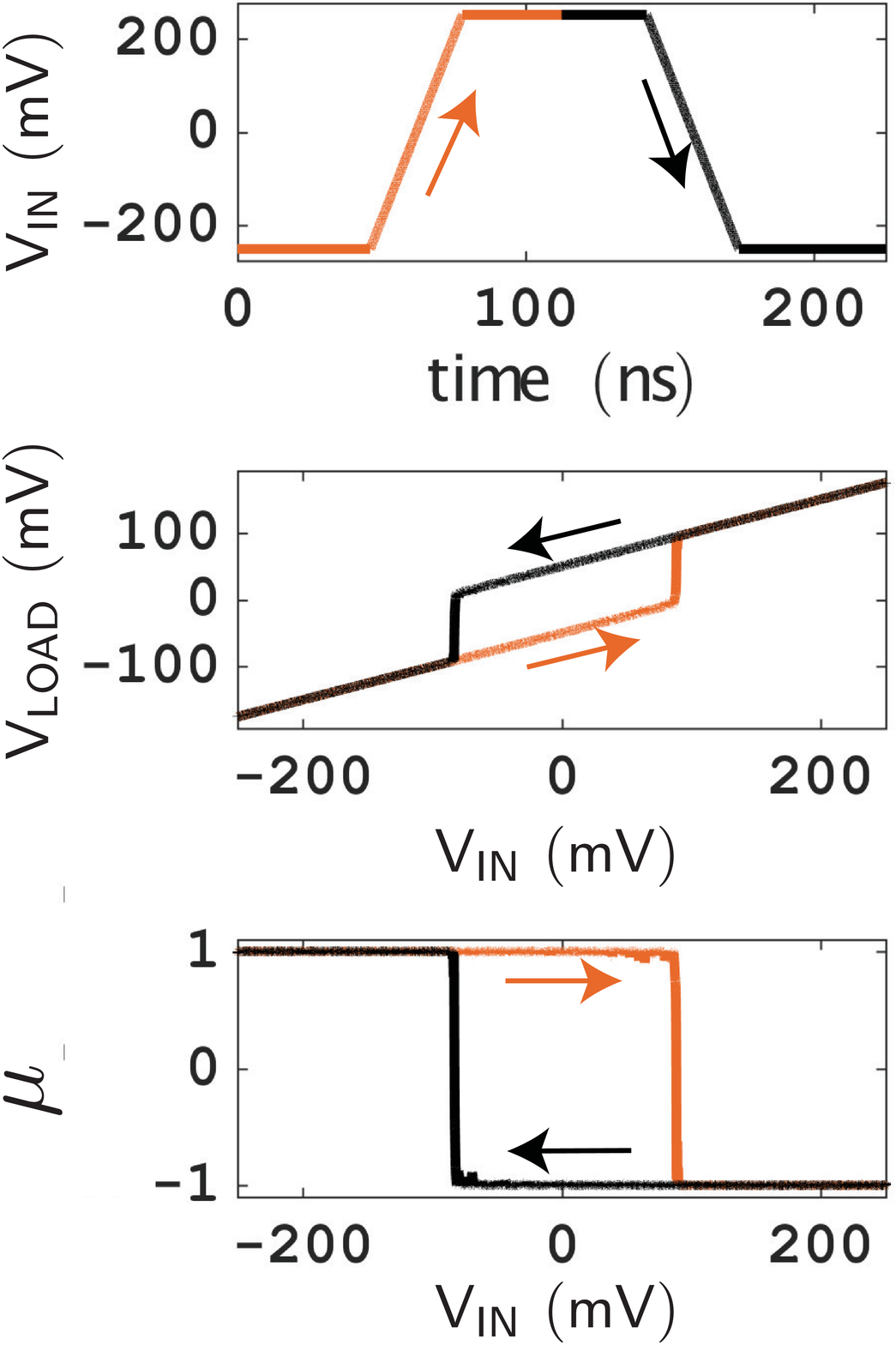}
\caption{\textbf{Non-volatility:} When ${C_{\rm eff}} v_M^2 /kT$ exceeds 1, the pseudo-magnetization ($\mu$) for the circular magnet ($H_K\rightarrow 0, E_A\rightarrow 0$) investigated in Fig.~\ref{fi:fig4} becomes stable. (a) Shows the input voltage ($\rm V_{IN}$) doing a negative-positive-negative sweep as a function of time. (b) Load voltage ($ C=C_L=100 $ aF  as a function of $\rm V_{IN}$. (c) $\mu$ as a function of $\rm V_{IN}$, exhibiting hysteresis.}
\label{fi:fig5}
\end{figure}

\section{Example \#2: Non-volatile operation}

It is easy to see by integrating Eq.~\ref{eq:blaw} that even when one uses a stable magnet ($E_A > 40\ kT$) in Eq.~\ref{eq:eq1}, the pseudo-magnetization $\mu$ does not show ``hysteretic'' behavior as function of $\rm V_{IN}$, but simply shifts the sigmoid response of Fig.~\ref{fi:fig4} to the left or right depending on the sign of $E_A$. The average sigmoidal behavior of Fig.~\ref{fi:fig4} is not just a consequence of using circular magnets, even a 40 kT magnet would show non-hysteretic behavior, but with suppressed fluctuations in $\mu$ and a shift along the $\rm V_{IN}$ axis. To obtain hysteretic behavior for the pseudo-magnetization $\mu$ we need an energy term that is quadratic ($\sim  \mu^2$) rather than linear ($=  E_A \ \mu$) as in Eq.~\ref{eq:eq1}. but we will not discuss the possibility further in this paper. Next we show that  the energy expression we have used could lead to hysteretic behavior if the ME coefficient $v_M$ were large enough. Such a quadratic term could arise naturally from the physics which we hope motivates future investigation. 

Fig.~\ref{fi:fig5} shows the results of a transient simulation of the equivalent circuit with a circular magnet, similar to Fig.~\ref{fi:fig4} with the only difference that in this example the back-voltage ($v_M$) is increased to 100 mV such that $C_{\rm eff} v_M^2 / kT \gg 1$. An input voltage is slowly swept from $-$200 mV to $+$200 mV and back to $-$200 mV within 1 $\mu$s, where pseudo-magnetization ($\mu)$ and the load voltage ($\rm V_L$) show hysteresis, similar to the magnetization of an ordinary magnet. One way to understand the hysteretic behavior is to note that the total energy for the full circuit in Fig.~\ref{fi:fig1} can be written as:
\begin{equation}
E_{total} = \frac{Q^2}{2 C_{\rm eff}} + Q v_M \mu -  Q V_{IN}
\end{equation}
\noindent where $C_{\rm eff}^{-1} = C^{-1} + C_L^{-1}$. 

Expanding Eq.~\ref{eq:bessel} for small $v_M$, we can approximate the pseudo-magnetization by $\mu\approx -{Q v_M}/{(2 kT)}$ and we have:
\begin{equation}
E_{total} \approx \frac{Q^2}{2 C_{\rm eff}}  - \frac{Q^2 v_M^2}{2 kT}  - Q V_{IN}
\end{equation}
\noindent suggesting that the ME effect provides a \textit{negative capacitance} $- kT / v_M^2 $ in series with $C_{\rm eff} $ leading to hysteretic behavior when $C_{\rm eff} v_M^2 > kT $ reminiscent of similar behavior based on ferroelectrics \cite{salahuddin2008use,khan2015negative}.

Numerical simulations of the equilibrium fluctuations of this magnet also show that the thermal stability of the $\mu$ is $\approx C_{\rm eff} v_M^2 / kT$ which can be 60 or greater, for reasonable values of $v_M$ and $C_{\rm eff}$ providing the possibility of non-volatile memory applications based on the pseudo-magnetization $\mu$.
\section{CONCLUSION}
In summary, we have presented  an equivalent circuit for magneotoelectric read and write and showed that it describes recent experiments on the MELRAM device quite accurately. We then used this circuit model to illustrate the possibility of representing ``1'' with different easy axes, encoded by the pseudo-magnetization $\mu$, rather than with different magnetizations, allowing a natural field-free operation that can be useful for a number of applications in stochastic neuromorphic computing. Lastly, we showed the possibility of using the pseudo-magnetization for non-volatile memory applications.
{\section*{ACKNOWLEDGMENT}}
The authors acknowledge insightful discussions with V. Ostwal, P.  Debashis, Z. Chen, J. Appenzeller, S. Majetich and J. T. Heron. 
This work was supported in part by the National Science Foundation through the NCN-NEEDS program, contract 1227020-EEC, the Nanoelectronics Research Initiative through the Institute for Nanoelectronics Discovery and Exploration (INDEX). 

\end{document}